\pdfoutput=1

\documentclass{article}

\newcommand{\mbx}{\mathbf{x}}
\newcommand{\mby}{\mathbf{y}}
\newcommand{\mbm}{\mathbf{m}}

\newcommand{\mbX}{\mathbf{X}}
\newcommand{\mbY}{\mathbf{Y}}
\newcommand{\mbM}{\mathbf{M}}

\newcommand{\mbv}{\mathbf{v}}
\newcommand{\mbw}{\mathbf{w}}

\newcommand{\bmt}{\bm{\theta}}
\newcommand{\bmT}{\bm{\Theta}}

\usepackage{setspace}
\usepackage{graphicx} 
\usepackage{subcaption}
\usepackage{cite}
\usepackage{multirow}
\usepackage{comment}
\usepackage{float}
\usepackage{amsmath,amssymb}
\usepackage{bm}
\usepackage[margin=1in]{geometry}

\usepackage{changepage}

\newlength\savedwidth

\newcommand\thickhline{\noalign{\global\savedwidth\arrayrulewidth\global\arrayrulewidth 2pt}%
\hline
\noalign{\global\arrayrulewidth\savedwidth}}

\usepackage{array}

\newcolumntype{+}{!{\vrule width 2pt}}

\title{Improved Disease Outbreak Detection from Out-of-sequence measurements Using Markov-switching Fixed-lag Particle Filters}

\author{Conor Rosato\textsuperscript{1,\ddag, *},
Joshua Murphy\textsuperscript{2, \ddag },
Si\^an E. Jenkins\textsuperscript{3},
Paul Horridge\textsuperscript{2}, \\
Alessandro Varsi\textsuperscript{2},
Martyn Bull\textsuperscript{3},
Alessandro Gerada\textsuperscript{1},
Alex Howard\textsuperscript{1}, \\
Veronica Bowman\textsuperscript{4} and 
Simon Maskell\textsuperscript{2}}
\date{%
\small{
    \textbf{1} Department of Pharmacology and Therapeutics, University of Liverpool, United Kingdom
\\
\textbf{2} Department of Electrical Engineering and Electronics, University of Liverpool, United Kingdom
\\
\textbf{3} RiskAware, United Kingdom
\\
\textbf{4} Paebbl Ltd
\\ [2ex]
    \ddag Equal Contribution\\
    * cmrosa@liverpool.ac.uk
}}

\begin{document}

\maketitle

\section*{Abstract}

Particle filters (PFs) have become an essential tool for disease surveillance, as they can estimate hidden epidemic states in nonlinear and non-Gaussian models. In epidemic modelling, population dynamics may be governed by distinct regimes such as endemic or outbreak phases which can be represented using Markov-switching state-space models. In many real-world surveillance systems, data often arrives with delays or in the wrong temporal order, producing out-of-sequence (OOS) measurements that pertain to past time points rather than the current one. While existing PF methods can incorporate OOS measurements through particle reweighting, these approaches are limited in their ability to fully adjust past latent trajectories. To address this, we introduce a Markov-switching fixed-lag particle filter (FL-PF) that resimulates particle trajectories within a user-specified lag window, allowing OOS measurements to retroactively update both state and model estimates. By explicitly reevaluating historical samples, the FL-PF improves the accuracy and timeliness of outbreak detection and reduces false alarms. We also show how to compute the log-likelihood within the FL-PF framework, enabling parameter estimation using Sequential Monte Carlo squared (SMC$^2$). Together, these contributions extend the applicability of PFs to surveillance systems where retrospective data are common, offering a more robust framework for monitoring disease outbreaks and parameter inference.

\section*{Author summary}

Early detection of disease outbreaks is essential for minimising both health and societal impacts. Our goal is to develop a statistical framework that serves as an early warning system for identifying disease outbreaks by fusing multiple streams of surveillance measurement data. In disease monitoring, data might be delayed, irregularly spaced, or arrive in OOS order, leading to challenges in timely analysis and response. We propose to use a FL-PF to provide daily estimates of the probability of the presence of an outbreak based on these asynchronous data. We show that through extensive simulation experiments, when compared with a standard PF, the FL-PF can provide more accurate estimates with fewer false-alarms. Additionally, our framework allows for efficient parameter estimation of the models with SMC$^2$. The code for this paper can be found https://github.com/j-j-murphy/FLEPI.

\section{Introduction}\label{sec:intro}
Syndromic surveillance is a method of monitoring public health to detect and respond to potential outbreaks of diseases \cite{fricker2013introduction}. It involves the real-time collection, analysis, and interpretation of health-related data to identify unusual patterns or clusters of symptoms that may indicate an outbreak. Detecting disease outbreaks in a timely manner is essential for minimising health impacts, protecting vulnerable populations, efficiently allocating resources, and ensuring economic stability. In particular, early warning systems can be effective in the detection of infectious diseases outbreaks \cite{meckawy2022effectiveness}.

A popular approach for outbreak detection is to employ Bayesian statistical principles to model and infer the likelihood of an outbreak occurring based on observed data. These methods provide a probabilistic framework that can incorporate prior information and update beliefs as new data becomes available. Some examples of Bayesian modelling for outbreak detection include: Bayesian network modelling \cite{jiang2010bayesian, wong2003bayesian, cooper2015method, cooper2012bayesian}, Bayesian Markov switching modelling \cite{lim2020inference, lin2014sequential, lu2009prospective}, Markov Chain Monte Carlo (MCMC) \cite{manitz2013bayesian, lim2020inference, lin2014sequential, lu2009prospective, adeoye2025bayesian} and Sequential Monte Carlo (SMC) methods \cite{dawson2015detecting, lin2014sequential}. 

Continually monitoring syndromic surveillance allows public health officials to predict changes in disease incidence patterns. These changes can result from shifts in model parameters, such as the transmissibility parameter, which can lead to abrupt increases in infection rates over time \cite{liu2024rtestim, morris2021inference, gill2023bayesian, murphy2024parameterizing} or decreases when disease control measures are being analysed \cite{feng2024switching}. One effective method for detecting changes in underlying incidence is the use of Markov switching models (MSM). MSMs have been successfully employed to detect outbreaks of diseases such as dengue \cite{lim2020inference}, influenza \cite{lin2014sequential, martinez2008bayesian, amoros2020spatio}, and anthrax \cite{lu2008bioterrorism}. The key idea is to capture the different phases or regimes of disease outbreaks, such as periods of low incidence (background noise) and high incidence (outbreaks), and to switch between these regimes based on the data observed over time. Predictions of the probability that there is an outbreak today (time $t$) can be made and evaluated using sensitivity (True Positive Rate), specificity (True Negative Rate) and the false alarm rate \cite{dawson2015detecting, shmueli2010statistical, lin2014sequential, martinez2008bayesian, zamba2008sequential, amoros2020spatio, cooper2012bayesian}.

\textit{Out-of-sequence} (OOS) syndromic measurements refer to data that is collected, reported, or obtained in a non-sequential or asynchronous manner, often arising in disease modelling when data collection is irregular, delayed, or reported without following a strict chronological order \cite{lee2016combining}. A recent review article outlines different nowcasting methods for accounting for reorting delays \cite{williams2024review}. Detecting disease outbreaks with OOS measurements can be challenging due to the non-linear nature of the data arrival \cite{shmueli2010statistical, johnson2025baseline}. Consequently, making inferences about disease states using data from various disparate sources can pose significant challenges \cite{de2015four}. Studies such as \cite{moore2022refining, rosato2023extracting, maishman2022statistical} combine heterogeneous data sources (e.g., death counts, hospital admissions, self-reported symptoms, positive test results, or social media signals) to infer the COVID-19 reproductive number $R_t$. These examples highlight how such data differ in spatio-temporal resolution and reporting latency. Latency in this context can refer to the delays between exposure, symptom onset, and infectiousness, often modeled by parameterising transmission kernels to extend the infectious period beyond a single time step \cite{li2018fitting}, or reporting delays between when an event occurs and when it is recorded. For example, self-reported symptom tweets are available daily with negligible delay \cite{rosato2023extracting}, whereas hospital admission counts are likely to be available with a delay. Uncertainty in these latency periods is commonly addressed using Bayesian inference methods \cite{moore2022refining, bastos2019modelling, birrell2021real}.

Particle filters (PFs) \cite{arulampalam2002tutorial} are online Bayesian methods that have been widely used to estimate the state of a disease based on noisy observations \cite{rasmussen2014phylodynamic, zimmer2017likelihood, dawson2015detecting, calvetti2021bayesian, rosato2022inference, endo2019introduction, kreuger2015particle, abm}. Traditionally, as data becomes available at each increment of time $t$, PFs provide an estimate of the probability distribution over the states of a disease model. These states could coincide with the latent compartments of the Susceptible, Infected, and Recovered (SIR) disease compartmental model \cite{kermack1927contribution}. As more data becomes available, this estimate is recursively updated. PFs can handle Markovian models wherein the future state of the system depends only on the current state and not on the sequence of events that preceded it. In the context of disease modeling, this means that the probability of an individual transitioning from one health state to another depends only on their current health status and not on how long they have been in that state or any prior disease history. These transition probabilities that govern the disease model can change over time based on Markov changes \cite{el2021particle}.

\begin{figure}[!h]
\centering
    \includegraphics[width=0.6\linewidth]{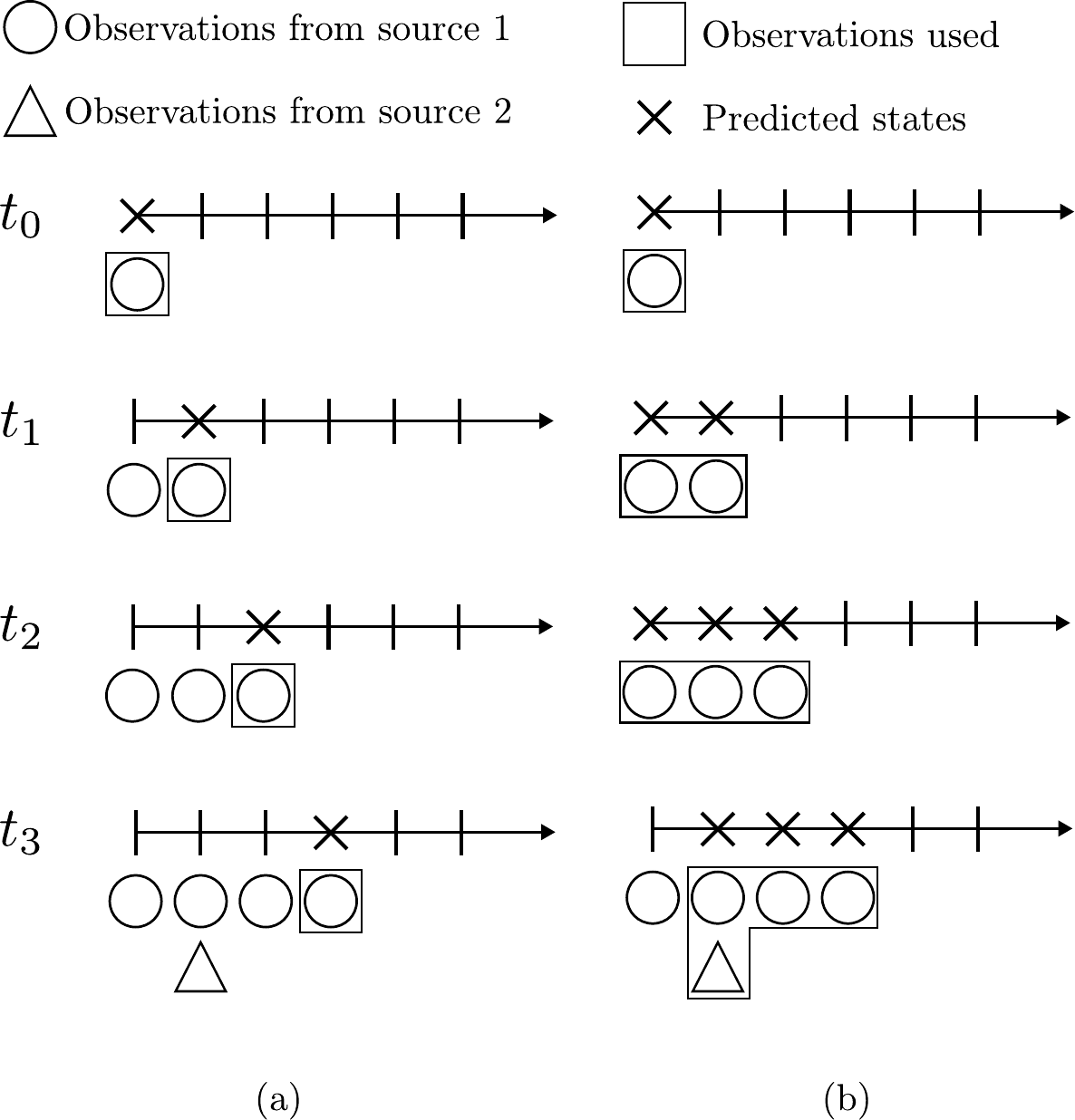}
    \caption{{\bf Graphical representation of how a PF and FL-PF analyse two incoming streams of data.}
    At times $t=0, 1, 2$ measurements are obtained from source 1 (circle). At $t=3$, an \textit{out-of-sequence} measurement which corresponds to $t=1$ from source 2 (triangle) is obtained. The rectangle outlines the measurements that are processed by the (a) PF and (b) FL-PF with lag 2.}
    \label{fig:flpf_images}
\end{figure}

As discussed, a delay in obtaining measurement data may exist. Standard PFs do not have the ability to adjust state estimates based on this OOS measurement due to their Markovian property (see Fig.~\ref{fig:flpf_images}). Previous work has looked at altering the PF to account for these OOS measurements by reweighting the particles when they are detected \cite{orton2001bayesian,orton2005particle,zhang2012out, maskell2006multi}. However, these methods don't re-evaluate state estimates in light of these OOS measurements which is essential when making timely outbreak predictions.

In this paper, we propose to use a Fixed-Lag Particle Filter (FL-PF) \cite{doucet2004fixed, doucet2006efficient, ruzayqat2022lagged, briers2006fixed, FL_NUTS}, also known as block sampling PFs \cite{johansen2015blocks}, to overcome the issue of analysing OOS measurements. At each time step, the FL-PF incorporates information from both the current and the previous $l$ observations that occur in the user specified lag window to estimate the state of the system (see Fig.~\ref{fig:flpf_images}). This allows the filter to account for delays in data reporting and adjust the state and dynamic model estimates accordingly. This is particularly advantageous when determining the presence of an outbreak, as data from several days prior can provide crucial indications. 

When modelling infectious diseases, it is also prudent to estimate the parameters of the model. For example, knowing the infection and recovery rates of a specific disease allows public health officials and researchers to ascertain the reproductive number $R_t$ at time $t$. This metric quantifies the average number of secondary infections produced by a single infected person. Particle-Markov Chain Monte Carlo (p-MCMC) \cite{andrieu_doucet_holenstein_2010} and Sequential Monte Carlo squared (SMC$^2$) \cite{chopin2013smc2} are two popular methods for parameter estimation when using a PF to model diseases \cite{rosato2022inference, smc2rosato, endo2019introduction, abm, calvetti2021bayesian}.

\section{Methods}
\subsection{Problem Statement}\label{sec:problem_statement}
Suppose we have a population of individuals, $N_{pop}$, each of which is assumed to belong to a particular health categorisation. These categories could include being Susceptible (S), Infected (I) or Recovered (R) from a disease \cite{kermack1927contribution}. Our aim is to track the number of people in each category on each day, given measurements of these unobserved categories from a number of sources. As outlined in the introduction, these sources may have a time delay, are ambiguous about the time they are referring to and/or are aggregated over a time interval, so a measurement can depend on the state history rather than just the state at the current time. On each day, a particular (unobserved) dynamic model is in effect which determines the evolution of the population state. Since these models could represent disease outbreaks, it is important to be able to estimate which model is in effect at each time-step.

\subsection{Markov-Switching State Space Model}\label{sec:ms_ssm}

The form of the problem statement above aligns with Markov-switching state space models (SSMs), as our goal is to infer both the latent states and regime dynamics from observed data within potentially non-linear and non-Gaussian systems.

MSMs are valuable tools for capturing sudden or gradual transitions between regimes within a system. These models are particularly useful for representing shifts in transmission dynamics, such as seasonal variations, public health interventions, or changes in pathogen virulence. A MSM operates under the assumption that disease dynamics alternate between different regimes, each defined by its own set of parameters.

The regimes are governed by a discrete-time Markov process, which switches between regimes according to transition probabilities. For a scenario where we are interested in whether the model is in an outbreak or non-outbreak regime, let $\mbM_{t}\in\{0,1\}$ represent the regime at time $t$. The probability of switching from regime 0 to regime 1 (or vice versa) is given by the transition matrix
    \begin{eqnarray}\label{eq:ms_transition_matrix}
        \mathcal{M} = \begin{bmatrix}
        p_{00} & p_{01} \\
        p_{11} & p_{11},
    \end{bmatrix}
    \end{eqnarray}
where $p_{00}$ is the probability of staying in a non-outbreak regime and $p_{01}$ is the probability of transitioning to an outbreak (similar definitions for $p_{10}$ and $p_{11}$).

An SSM is typically composed of a state equation and an observation equation, but a MSM also contains the transition matrix from Eq \eqref{eq:ms_transition_matrix} such that the complete formulation is described by  
    \begin{align}
        \mbM_t|\mbM_{t-1} &\sim \mathcal{M}^{\mbM_t\mbM_{t-1}},\\
         \mbX_{t} \mid \mbX_{t-1} &\sim f(\mbX_{t} \mid \mbX_{t-1}, \bmT_{\mbM_t}),\label{xt}\\
        \mbY_t \mid \mbX_t &\sim h(\mbY_t \mid \mbX_t, \bmT_{\mbM_t}).
    \end{align}

The two arbitrary functions, $f_{\mbM_t}$, and $h_{\mbM_t}$, which correspond to the state and observation equations, respectively, can be nonlinear and non-Gaussian. The SSM is parameterised by $\bm{\theta} \in \mathbb{R}^{D_{\bmt}}$ but note the dependence of the parameters $\bmt_{\mbM_t}$ in the state and observation equation on the regime that the SSM is in. An SSM sequentially simulates the latent states, $\mbX_{1:T}=\{\mbX_1,\dots, \mbX_t, \dots, \mbX_T\}$, conditional on the regime $\mbM_{1:T}=\{\mbM_1,\dots, \mbM_t, \dots, \mbM_T\}$ and a stream of incoming measurements received at each time-step $t$, $\mbY_{1:T}=\{\mbY_1,\dots, \mbY_t, \dots, \mbY_T\}$, over $T$ time-steps. The joint distribution at $t$ is 
\begin{align}
    p(\mbX_{1:t},\mbM_{1:t}, \mbY_{1:t}|\bmT)=p(\mbM_1)f(\mbX_1&|\bmT)h(\mbY_1|\mbX_1,\bmT)\times \nonumber\\
    &\prod_{\tau=2}^t \mathcal{M}^{\mbM_t\mbM_{t-1}}f(\mbX_\tau|\mbX_{\tau-1}\bmT)h(\mbY_\tau|\mbX_\tau,\bmT).\label{eq:ssm_joint}
\end{align}

\subsection{Particle Filter}\label{sec:particle_filter}

\begin{figure}[!h]
    \includegraphics[width=0.8\linewidth]{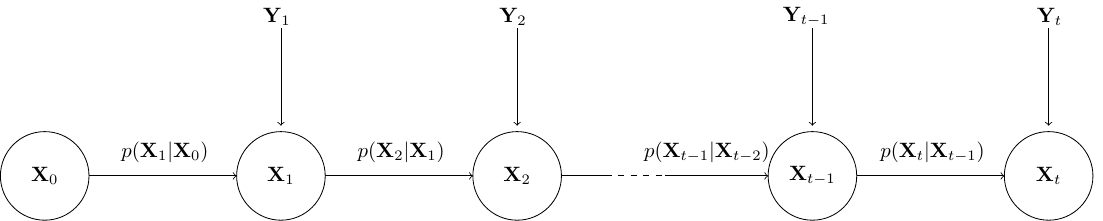}
    \caption{{\bf PF: state flow}}
    \label{fig:PF_flow}
\end{figure}

The joint distribution of an SSM in Eq \ref{eq:ssm_joint} is typically intractable but can be estimated using PFs (see Fig.~\ref{fig:PF_flow}). PFs are an SMC method which use a combination of sequential importance sampling (SIS) and resampling steps to make estimates of the dynamic true state $\mbX_t \in \mathbb{R}^{D_\mbX}$ and model $\mbM_t \in \mathbb{Z}^{1}$ using a set of $N_x$ particles which contain hypotheses of the state $\mbx_t \in \mathbb{R}^{N_x \times D_\mbX}$ and model $\mbm_t \in \mathbb{R}^{N_x}$ with weights $\mbv_{1:t} \in \mathbb{R}^{N_x}$. 

At time-step $t = 1$, the particles are initialised using a proposal distribution, $q_1(\cdot)$
\begin{align}
    (\mathbf{x}_{1}^{j}, \mbm_{1}^{j}) &\sim q_1(\cdot), \ \ \ \forall j,
\end{align}
and weighted uniformly
\begin{align}
    \mbv_0^{j} &= \frac{1}{N_x}, \ \ \ \forall j.
\end{align}

As time evolves, SIS draws new particles from a proposal distribution, $q(\mbx_{t}, \mbm_t|\mbx_{t-1}, \mbm_{t-1}, \bm{\theta}, \mbY_{t})$
     \begin{equation}\label{eq:pf_proposal}
         (\mathbf{x}_{t}^{i}, \mbm_{t}^{i}) \sim q(\cdot|\mbx_{t-1}, \mbm_{t-1}, \bm{\theta}, \mbY_{t}),
    \end{equation}
and weighted by
\begin{align}
\mbv_{1:t}^{j}= \mbv_{1:t-1}^{j}\frac{p\left(\mbY_t|\mathbf{x}_t^{j}, \mbm_t^j,\bm{\theta}\right)p\left(\mathbf{x}_t^{j}, \mbm_t^j,|\mathbf{x}^{j}_{t-1},\mbm_{t-1}^j,\bm{\theta}\right)}{q\left(\mathbf{x}_t^{j}, \mbm_t^j,|\mathbf{x}_{t-1}^{j}, \mbm_{t-1}^j, \bm{\theta},\mbY_t\right)}. \ \ \ \forall j.\label{eq:pf_weightupdate}
\end{align}

After performing the SIS step, the weights are normalised by
    \begin{equation}
        \tilde{\mbv}_{1:t}^{j} = \frac{\mbv_{1:t}^{j}}{\sum_{j'=1}^{N_x}\mbv_{1:t}^{j'}}, \ \ \ \forall j.
    \end{equation}

which ensures the weights sum to $1$, i.e., $100\%$ of the total probability space. At this point, it is possible to generate unbiased estimates of the true state and model through
    \begin{align} \label{eq: estimate_pf}
        \hat{X}_t = \sum_{j=1}^{N_x} \mathbf{\tilde{\mbv}}_{1:t}^{j}\mbx_t^{j},\\
        \hat{M}_t = \sum_{j=1}^{N_x} \mathbf{\tilde{\mbv}}_{1:t}^{j}\mbm_t^{j},
    \end{align} respectively. 

As time evolves, particle degeneracy occurs such that all the normalised weights bar one tend to $0$, while a single weight tends towards $1$, resulting in the estimates in Eq~(\ref{eq: estimate_pf}) becoming less useful. To address this issue, the Effective Sample Size (ESS$_x$)
    \begin{equation}
        \text{ESS}_x = \frac{1}{\sum_{j=1}^{N_x} (\mathbf{\tilde{\mbv}}_t^{j})^2},
    \end{equation}
is used, with resampling triggered when ESS$_x$ falls below a chosen threshold, usually $N_x / 2$. Resampling generates a new population of particles by drawing from the current particle set $N_x$ times, proportional to the associated weights of the particles. Multiple resampling algorithms exist \cite{resamplingmethods} but in this paper we employ multinomial resampling within the PF \cite{smc2rosato}.

\subsubsection{Model and State propagation}\label{sec:model_propogation}

Suppose the proposal distribution is chosen such that it can be factorised as 
    \begin{align}
        q(\mbx_{t}, \mbm_t|\mbx_{t-1}, \mbm_{t-1}, \bm{\theta}, \mbY_{t})&=q(\mathbf{x}_{t} \mid \mathbf{x}_{t-1}, \mbm_{t})q(\mbm_t|\mbm_{t-1})\\
        &=q(\mathbf{x}_{t} \mid \mathbf{x}_{t-1}, \mbm_{t})\mathcal{Q}^{\mbm_{t-1},\mbm_{t}},
    \end{align}
 where $\mathcal{Q}$ is a Markov transition matrix. The dynamics and proposal components of Eq~\eqref{eq:pf_weightupdate} are defined as 
 
    \begin{align}
       \frac{f(\mathbf{x}_{t}\mid\mathbf{x}_{t-1}, \bmt_{\mbm_t})\mathcal{M}^{\mbm_{t-1}, \mbm_t}}{q(\mbx_{t}, \mbm_t|\mbx_{t-1}, \mbm_{t-1}, \bm{\theta}, \mbY_{t})}=\frac{f(\mathbf{x}_{t}\mid\mathbf{x}_{t-1}, \bmt_{\mbm_t})\mathcal{M}^{\mbm_{t-1}, \mbm_t}}{q(\mathbf{x}_{t} \mid \mathbf{x}_{t-1}, \mbm_{t})\mathcal{Q}^{\mbm_{t-1},\mbm_{t}}}\label{eq:model_definition}.
    \end{align}

In Eq~(\ref{eq:model_definition}), if $\mathcal{Q}_t=\mathcal{M}_t$ and $q(\mathbf{x}_{t} \mid \mathbf{x}_{t-1}, \mbm_{t})=f(\mathbf{x}_{t}\mid\mathbf{x}_{t-1}, \bmt_{\mbm_t})$, then the fraction cancels to leave a PF weight update as
\begin{equation}\label{eq:pf_weight_update_dynamics_prop}
    \mbv_{1:t}^{j}= \mbv_{1:t-1}^{j}p\left(\mbY_t|\mathbf{x}_t^{j}, \mbm_t^j,\bm{\theta}\right), \ \ \ \forall j.
\end{equation}

However, if certain model switches are rare (such as those modelling disease outbreak), it can be beneficial to over-sample them in order to ensure that sufficient number of particles within the PF represent these models.

\subsubsection{Likelihood with OOS Measurements}

There can be multiple sensors $S$, observing the latent state such that the dimensionality could be up to $\mbY_t \in \mathbb{R}^{S\times D_{\mbY}}$, if all sensors take measurements at that time-step. There may also be a disparity between the time-step when a measurement is generated $t_g$ and when a measurement is received $t_r$ such that a measurement is OOS. A single measurement from a sensor $s$, generated at $t_g$ and received at $t_r$ is thus denoted by $\mby_{t_g,t_r}^s$. The likelihood $p\left(\mbY_t|\mathbf{x}_t^{j}, \mbm_t^j,\bm{\theta}\right)$ in Eq~\ref{eq:pf_weightupdate} and Eq~\ref{eq:pf_weight_update_dynamics_prop} can then be calculated from
\begin{equation}\label{eq:pf_likelihood_OOS}
    p\left(\mbY_t|\mathbf{x}_t^{j}, \mbm_t^j,\bm{\theta}\right) = \prod_{s=1}^S h(\mby_{t_g,t_r}^s| \mathbf{x}_t^{j}, \bm{\theta}_{\mbm_t^j})^{\mathbb{I}(t_g=t)\cap\mathbb{I}(t_r=t)}, \ \ \ \forall j,
\end{equation}
where $\mathbb{I}(\cdot)$ is an indicator function. This means that a measurement will contribute to the likelihood and thus particle weight if its time of generation and reception are the same as the current time-step. This means a standard PF configuration will not account for OOS measurements. However, there are PF algorithms which will account for OOS measurements by recursively modifying the PF weights from the time-step at which the measurement was generated but not altering the PF state estimates \cite{orton2001bayesian,orton2005particle,zhang2012out}.

\subsection{Fixed-lag Particle Filter}
The FL-PF extends the PF by reprocessing historic measurements at previous time steps which can lead to more accurate approximations of the posterior distribution over the state. Each particle within the FL-PF is no longer a hypothesis of the most recent true state, $\mathbf{X}_t$, but a hypothesis of a block (or trajectory) of $l+1$ states $\mathbf{X}_{t-l:t}$, where $l \in \mathbb{N}$ is a lag selected by the user. More precisely, at each time step $t$, the FL-PF proposes each new particle, $\mathbf{x}_{t-l:t}^{j}$ $\forall j = 1, 2, \dots, N_x$, by reverting back to time step $t-l-1$, and (re)sampling all the states for the time steps, $\tau = t-l, t-l+1, \dots, t$, given $\mathbf{x}_{t-l-1}^{j}$ and the most recent $l+1$ measurements, $\mathbf{Y}_{t-l:t}$, which are stored in memory. Therefore, the previous $l$ states of each particle (which we denote as $\mathbf{\overline{x}}_{t-l:t-1}^{j}$) are corrected, and the hypothesis on the most recent state $\mathbf{x}_t^{j}$ will be a better approximation of $\mathbf{X}_t$. 

\begin{figure} [!ht]
    \centering
    \includegraphics[width=0.8\linewidth]{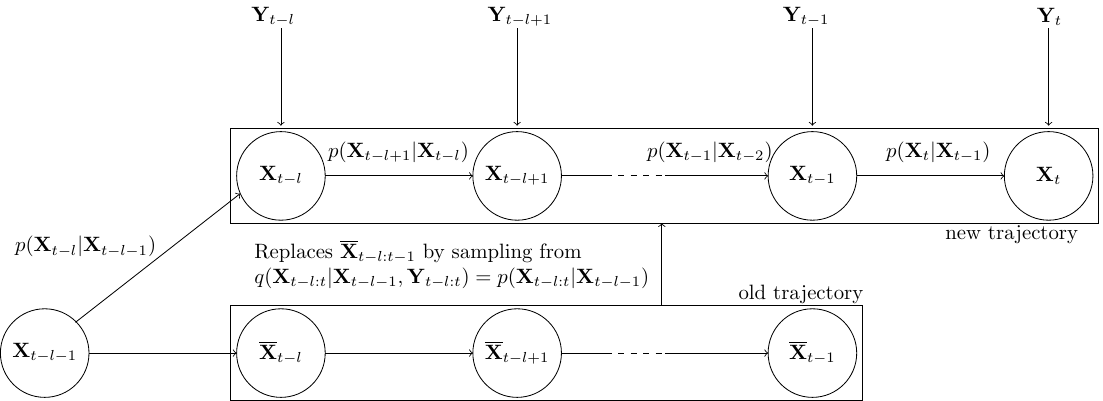}
    \caption{FL-PF: state flow} \label{fig: fl_pf}
\end{figure}

Following \eqref{eq:pf_proposal}, the proposal of an FL-PF can be denoted 
\begin{align}\label{eq:flpf_proposal}
    (\mathbf{x}_{t-l:t}^{j}, \mbm_{t-l:t}^{j}) &\sim q(\cdot\mid\mathbf{x}_{t-l-1}^{j}, \mbm_{t-l-1}^{j}, \bmt, \mathbf{Y}_{t-l:t}),  \ \ \ \forall j
\end{align}
with a corresponding weight update given by 
\begin{align}\label{eq:flpf_weightupdate}
    \mbv_{1:t}^{j} = \mbv_{1:t-1}^{j} &\frac{\pi(\mathbf{x}_{t-l:t}^{j}, \mathbf{m}_{t-l:t}^{j}, \mathbf{Y}_{t-l:t} \mid \mathbf{x}_{t-l-1}^{j}, \mathbf{m}_{t-l-1}^{j}, \bm{\theta})}{\pi(\mathbf{\overline{x}}_{t-l:t-1}^{j}, \mathbf{\overline{m}}_{t-l:t-1}^{j}, \mathbf{Y}_{t-l:t} | \mathbf{x}_{t-l-1}^{j},\mathbf{m}_{t-l-1}^{j}, \bm{\theta})}\times\nonumber\\
    &\frac{L(\mathbf{\overline{x}}_{t-l:t-1}^{j},\mathbf{\overline{m}}_{t-l:t-1}^{j} \mid \mathbf{x}_{t-l:t}^{j}, \mathbf{m}_{t-l:t}^{j})}{q(\mathbf{x}_{t-l:t}^{j}, \mathbf{m}_{t-l:t}^{j}\mid\mathbf{x}_{t-l-1}^{j}, \mathbf{m}_{t-l-1}^{j}, \bmt, \mathbf{Y}_{t-l:t})}, \ \ \ \forall j,
\end{align}
where
    \begin{align}
        \pi(\mathbf{x}_{t-l:t}^{j}, \mathbf{m}_{t-l:t}^{j}, \mathbf{Y}_{t-l:t} \mid &\mathbf{x}_{t-l-1}^{j}, \mathbf{m}_{t-l-1}^{j}, \bm{\theta}) = \nonumber\\ &\prod_{\tau = t-l}^{t} 
        p\left(\mbY_\tau|\mathbf{x}_\tau^{j}, \mbm_\tau^j,\bm{\theta}\right)p\left(\mathbf{x}_\tau^{j}, \mbm_\tau^j,|\mathbf{x}^{j}_{\tau-1},\mbm_{\tau-1}^j,\bm{\theta}\right), \ \ \ \forall j,
    \end{align}
is the joint incremental posterior that we aim to sample from and $L(\mathbf{\overline{x}}_{t-l:t-1}^{j},\mathbf{\overline{m}}_{t-l:t-1}^{j} \mid \mathbf{x}_{t-l:t-1}^{j}, \mathbf{m}_{t-l:t-1}^{j})$ is a user-designed backward kernel. In an FL-PF, we sample from the dynamics $p(\mathbf{X}_{t-l:t}\mid\mathbf{X}_{t-l-1})$, and the backward kernel is chosen to match the proposal, i.e., $L(\cdot) = q\left(\overline{\mathbf{x}}_{t-l:t-1}^{j}, \overline{\mathbf{m}}_{t-l:t-1}^{j} | \mathbf{x}^{j}_{t-l-1}, \mathbf{m}^{j}_{t-l-1}\right)$ (see Fig.~\ref{fig: fl_pf}). The FL-PF performs all steps described in Section \ref{sec:particle_filter} for the PF, but \eqref{eq:flpf_proposal} replaces \eqref{eq:pf_proposal} and \eqref{eq:flpf_weightupdate} replaces \eqref{eq:pf_weightupdate}, such that, in this configuration, a FL-PF with $l = 0$ is identical to a PF.

With a FL-PF filter, OOS measurements can be incorporated into the likelihood so long as the lag window contains both $t_g$ and $t_r$
\begin{equation}\label{eq:flpf_likelihood_OOS}
    p\left(\mbY_t|\mathbf{x}_t^{j}, \mbm_t^j,\bm{\theta}\right) = \prod_{s=1}^S h(\mby_{t_g,t_r}^s| \mathbf{x}_t^{j}, \bm{\theta}_{\mbm_t^j})^{\mathbb{I}(t_g=t)\cap\mathbb{I}(t_r\leq t+l)}.
\end{equation}

\subsection{Sequential Monte Carlo Squared for parameter estimation}\label{sec:smc_squared}
As explained in Section~\ref{sec:intro}, it is helpful to infer the parameters that govern the spread of a disease. In this paper, we use the SMC$^2$ framework outlined in \cite{smc2rosato} to estimate $\bm{\theta}$. The framework is parallelised using Message Passing Interface (MPI) which exploits parallel computing architectures. The resulting SMC$^2$ algorithm runs in optimal $\mathcal{O}(\log_2N)$ time complexity which is bounded by the parallel resampling algorithm outlined in \cite{Alessandro5, Alessandro6}. 

SMC$^2$ consists of two SMC methods: an SMC sampler and a PF. At the SMC sampler level, estimates are made of the static parameters $\bmT \in \mathbb{R}^{D_{\bmT}}$ over $K$ time-steps using a set of $N$ samples which contain hypotheses of the parameters $\bmt_k \in \mathbb{R}^{N\times D_{\bmT}}$ with weights $\mbw_k \in \mathbb{R}^N$. SMC$^2$ targets $\pi(\mathbf{\bm{\theta}})=q_1(\bm{\theta})p(\mathbf{Y}_{1:T}|\bm{\theta})$ via SIS steps and resampling \cite{chopin2013smc2}, where $q_1(\bm{\theta})$ is the prior and $p(\mathbf{Y}_{1:T}|\bm{\theta})$ is the likelihood, estimated by the PF or FL-PF. Each sample thus runs a PF with its hypothesised parameters $\bmt^i_k$ and calculates a likelihood as the sum of the weights of a PF at its final time-step
\begin{equation}
    p(\mathbf{Y}_{1:T}|\bm{\theta})=\frac{1}{N_x}\sum_{j=1}^{N_x} \mathbf{\mbv}_{1:T}
\end{equation}
The static joint distribution which SMC$^2$ targets at the SMC sampler level is 
     \begin{equation}
        \pi(\mathbf{\bm{\theta}}_{1:K}) = \pi(\mathbf{\bm{\theta}}_{K}) \prod_{k=2}^{K} L(\mathbf{\bm{\theta}}_{k-1} | \mathbf{\bm{\theta}}_{k}),
    \end{equation}
where $L(\mathbf{\bm{\theta}}_{k-1} | \mathbf{\bm{\theta}}_{k})$ is the L-kernel, which is a user-defined probability distribution. At $k=1$, $N$ samples $\forall i = 1, \dots, N$ are drawn from a prior distribution $q_1(\bm{\theta}^i_1)$ and weighted according to
    \begin{equation} 
        \mathbf{w}^i_1 = \frac{\pi(\mathbf{\bm{\theta}}^i_1)}{q_1(\mathbf{\bm{\theta}}^i_1)}, \ \ \forall i\label{init_weights}.
    \end{equation}
At $k>1$, subsequent samples are proposed via a proposal distribution, $q(\mathbf{\bm{\theta}}^i_k|\mathbf{\bm{\theta}}^i_{k-1})$, and weighted by 
    \begin{equation}
        \mathbf{w}^i_{k} = \mathbf{w}^i_{k-1} \frac{\pi(\mathbf{\bm{\theta}}^i_{k})}{\pi(\mathbf{\bm{\theta}}^i_{k-1})} \frac{L(\mathbf{\bm{\theta}}^i_{k-1}|\mathbf{\bm{\theta}}^i_{k})}{q(\mathbf{\bm{\theta}}^i_{k}|\mathbf{\bm{\theta}}^i_{k-1})}, \ \ \forall i.\label{eq:l_weights}
    \end{equation}
SMC sampler weights are normalised using
\begin{equation}
        \tilde{\mbw}_t^{j} = \frac{\mbw_t^{j}}{\sum_{i'=1}^{N}\mbw_t^{i'}}, \ \ \ \forall i.
    \end{equation}
As in a PF, resampling occurs if the effective sample size, at the SMC sampler level denoted by ESS$_{\bmt}$ and calculated by
\begin{equation}
        \text{ESS}_{\bmt} = \frac{1}{\sum_{i=1}^{N} (\mathbf{\tilde{\mbw}}_t^{i})^2},
    \end{equation}
falls below $\frac{N}{2}$. A parallelised form of systematic resampling is used at the SMC sampler level rather than multinomial resampling as in the PF \cite{Alessandro5, Alessandro6}.
    
Estimates of parameters can be made by
\begin{equation}
    \hat{\bmt} = \sum_{i=1}^{N} \tilde{\mathbf{w}}^i_{k}\bm{\theta}^i_{k}, \label{eq:realised_estimates_smc}
\end{equation}
with samples from previous iterations incorporated through recycling \cite{nguyen2015efficient}.


\subsection{SEIRS Model}\label{sec:SEIRS_state}

Returning to the problem statement in Section \ref{sec:problem_statement}, our interest in categorising health status in relation to a specific disease can be described by the stochastic Susceptible, Exposed, Infected, Recovered, Susceptible (SEIRS) compartmental disease model \cite{kermack1927contribution}. For clarity we note that the E and I compartments contain individuals that are exposed but not infectious and those that are infectious, respectively. In the SEIRS model, individuals are recirculating from being recovered to the same S department due to disease mutation, for example. We assume a single geographic location and that each of a fixed number of people $N_{pop}$ is in one of the following four compartments on the $t-$th day:
    \begin{enumerate}
    \item $\mbX_t^1=\mathbf{S}_t$: Susceptible - Individuals who are not infected but are at risk of being exposed to the disease.
    \item $\mbX_t^2=\mathbf{E}_t$: Exposed - Individuals who have been exposed to the disease and are assumed to be in the incubation period. They are not yet infectious but will become infectious after some time.
    \item $\mbX_t^3=\mathbf{I}_t$: Infected - Individuals who are currently infectious and can transmit the disease to susceptible individuals.
    \item $\mbX_t^4=\mathbf{R}_t$: Recovered - Individuals who have recovered from the disease and have developed immunity, no longer at risk of infection or spreading the disease.
    \end{enumerate}
%
The dimensionality of the discrete state is therefore $D_{\mbX}=4$. 

The state equation for an SEIRS model describes the number of people leaving each of the compartments at each time-step as a binomially distributed random variable. The binomial probabilities, $p(\cdot\rightarrow \cdot)$, can be regime $\mbM_t$ specific and are defined as the rate at which individuals leave a compartment. Individuals are assumed to transition between states independently. The corresponding binomial distributions for the SEIRS model are
    \begin{align}
    n(\mathbf{S}\rightarrow \mathbf{E})&\sim \text{Binomial}(\mathbf{S}_{t-1}, p(\mathbf{S}\rightarrow \mathbf{E})), \label{discrete_s_binom}\\
    n(\mathbf{E}\rightarrow \mathbf{I})&\sim \text{Binomial}(\mathbf{E}_{t-1}, p(\mathbf{E}\rightarrow \mathbf{I}))\label{discrete_e_binom},\\
    n(\mathbf{I}\rightarrow \mathbf{R})&\sim \text{Binomial}(\mathbf{I}_{t-1}, p(\mathbf{I}\rightarrow \mathbf{R}))\label{discrete_i_binom},\\
    n(\mathbf{R}\rightarrow \mathbf{S})&\sim \text{Binomial}(\mathbf{R}_{t-1}, p(\mathbf{R}\rightarrow \mathbf{S})),\label{discrete_r_binom}
    \end{align}
where $p(\mathbf{S}\rightarrow \mathbf{E})$, $p(\mathbf{E}\rightarrow \mathbf{I})$, $p(\mathbf{I}\rightarrow \mathbf{R})$ and $p(\mathbf{R}\rightarrow \mathbf{S})$ denote the probability of leaving the susceptible, exposed, infected and recovered compartments, respectively and are given by 
    \begin{align}
    p(\mathbf{S}\rightarrow \mathbf{E})&=1-e^{-\frac{\beta \mathbf{I}_{t-1}}{N_{pop}}}, \label{discrete_s_p}\\
    p(\mathbf{E}\rightarrow \mathbf{I})&=1-e^{-\gamma}\label{discrete_e_p},\\
    p(\mathbf{I}\rightarrow \mathbf{R})&=1-e^{-\sigma}\label{discrete_i_p},\\
    p(\mathbf{R}\rightarrow \mathbf{S})&=1-e^{-\xi}\label{discrete_r_p}.
    \end{align}
The total count of individuals moving compartments is governed by the parameters $\bmT =\lbrack\beta, \gamma, \sigma, \xi\rbrack$. The complete discrete, stochastic SEIRS model is therefore presented as
    \begin{align}
    \mathbf{S}_{t}&=\mathbf{S}_{t-1}-n(\mathbf{S}\rightarrow \mathbf{E})+n(\mathbf{R}\rightarrow \mathbf{S}), \label{SIRR:s_discrete}\\
    \mathbf{E}_{t}&=\mathbf{E}_{t-1}+ n(\mathbf{S}\rightarrow \mathbf{E}) - n(\mathbf{E}\rightarrow \mathbf{I}),\label{SIRR:e_discrete}\\
    \mathbf{I}_{t}&=\mathbf{I}_{t-1}+ n(\mathbf{E}\rightarrow \mathbf{I})-n(\mathbf{I}\rightarrow \mathbf{R})\label{SIRR:i_discrete},\\
     \mathbf{R}_{t}&=\mathbf{R}_{t-1}+ n(\mathbf{I}\rightarrow \mathbf{R})-n(\mathbf{R}\rightarrow \mathbf{S}).\label{SIRR:r_discrete}
    \end{align} 

\begin{figure}[!h]
\centering
    \includegraphics[width=0.5\linewidth]{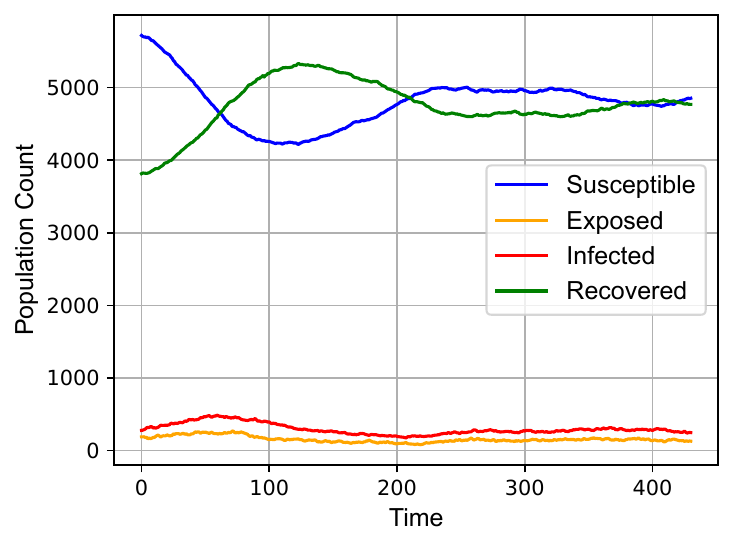}
    \caption{{\bf SEIRS model trajectories with population 10{,}000 over 430 days.}
    The Susceptible (blue), Exposed (orange), Infected (red), and Recovered (green) model outlined in \protect{Section~\ref{sec:SEIRS_state}} simulated for 430 days with a population of 10{,}000. The model parameters are $\bm{\theta} = [\beta, \gamma, \sigma, \delta] = [0.2, 1/5, 1/10, 1/180]$.}
    \label{fig:seirs_images}
\end{figure}

An example of the SEIRS model trajectories with population$=10000$ over 430 days can be seen in Fig.(\ref{fig:seirs_images}). 

\subsubsection{SEIRS Markov-switching model}\label{sec:SEIRS_MS}
During an outbreak, the transmissibility of a disease is higher than in a non-outbreak period. The SEIRS disease model in Section~\ref{sec:SEIRS_state} can be expanded to reflect the fact that $\beta(t)$ depends on the regime $\mbM_t$. This would change the probability of transition between compartments in Eq~(\ref{discrete_s_p}) to be
    \begin{eqnarray}
         p(\mathbf{S}\rightarrow \mathbf{E})&=\frac{\beta_{\mbM_t} \mathbf{I}_{t-1}}{N_{pop}},
    \end{eqnarray}
where $\mbM_t$ determines whether the system is in the high or low transmission regime. The set of parameters govering the SEIRS model is now $\bmT =\lbrack\beta^0, \beta^1, \gamma, \sigma, \xi\rbrack$.

\begin{figure*}[!ht]
\centering
\begin{subfigure}[t]{0.48\linewidth}
    \centering
    \includegraphics[width=\linewidth]{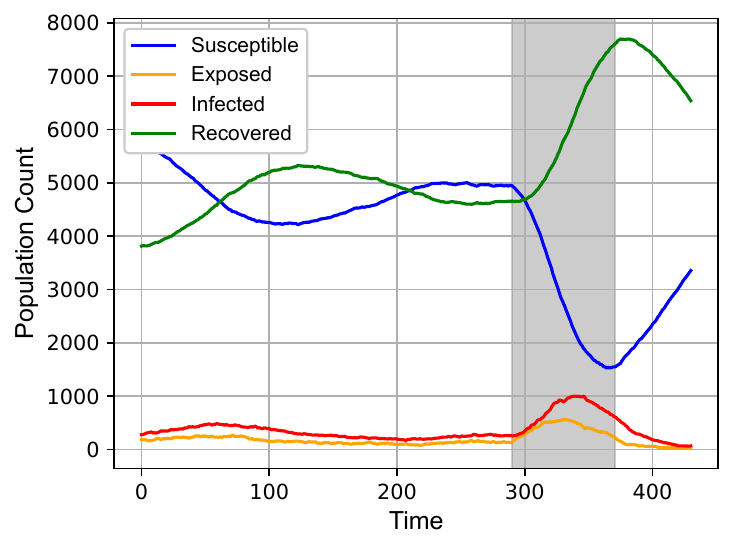}
    \caption{}
    \label{fig:outbreak_1_8}
\end{subfigure}
\hfill
\begin{subfigure}[t]{0.48\linewidth}
    \centering
    \includegraphics[width=\linewidth]{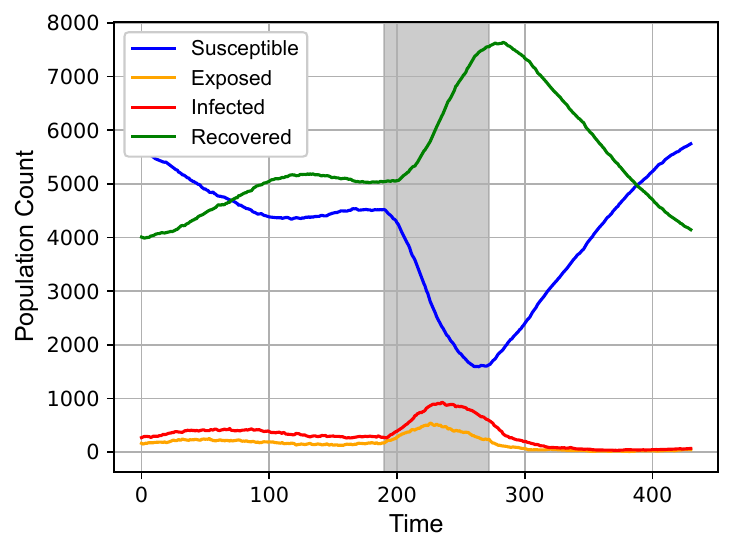}
    \caption{}
    \label{fig:outbreak_1_7}
\end{subfigure}

\vspace{0.5em} 

\begin{subfigure}[t]{0.48\linewidth}
    \centering
    \includegraphics[width=\linewidth]{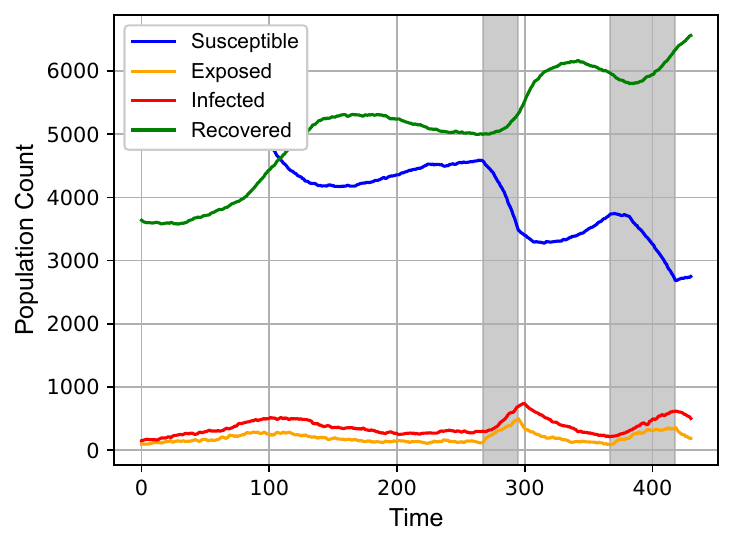}
    \caption{}
    \label{fig:outbreak_50_2}
\end{subfigure}
\hfill
\begin{subfigure}[t]{0.48\linewidth}
    \centering
    \includegraphics[width=\linewidth]{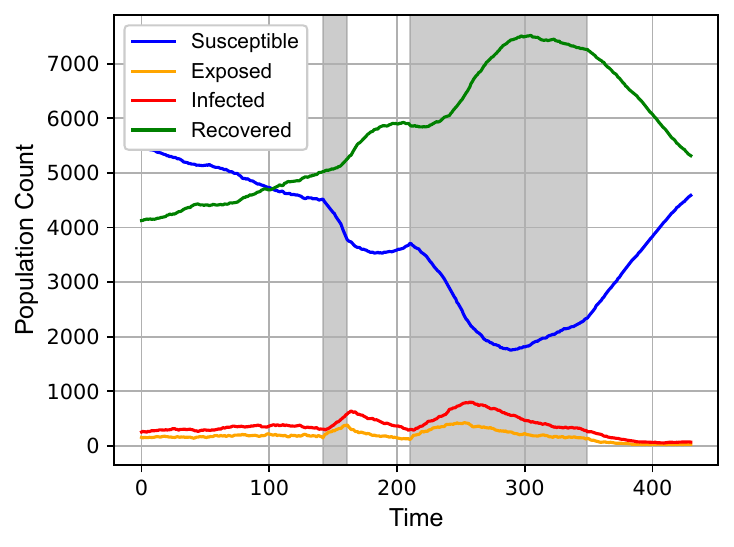}
    \caption{}
    \label{fig:outbreak_50_5}
\end{subfigure}

\caption{{\bf SEIRS model trajectories with population 10000 over 430 days with randomly generated outbreaks.}
    The Susceptible (blue), Exposed (orange), Infected (red), and Recovered (green) compartments from the SEIRS model (\protect{Section~\ref{sec:SEIRS_state}}) simulated for 430 days. A population of 10000 is used, and the first 200 days are discarded to allow the system to reach steady state. Outbreaks correspond to the gray shaded time periods. Model parameters during non-outbreak periods are $\bm{\theta}_{\mbM^0} = [\beta, \gamma, \sigma, \xi] = [0.2, 1/5, 1/10, 1/180]$. During outbreaks, $\beta$ increases to 0.4. Subplots (A) and (B) show single-outbreak simulations; (C) and (D) show simulations with two outbreaks.}
\label{fig:simulated_outbreaks}
\end{figure*}

Fig.~\ref{fig:simulated_outbreaks} outlines examples of one and two randomly generated outbreaks presented on the top and bottom rows, respectively. During an outbreak period, $\beta$ changes from 0.2 to 0.4 resulting in more people being exposed to the disease and then becoming infected. 

\subsubsection{SEIRS Measurement Model}\label{sec:SEIRS_likelihood}
This section outlines the Poisson measurement model, which noisy count of individuals in the infected compartment. 



For a sensor index $s$, $\mby_{t_g,t_r}^s \in \mathbb{Z}^{+}$ is the number of observed infections at $t_g$ and reported at $t_r$. The observation equation for that sensor is then given by
    \begin{eqnarray}
         h(\mathbf{y} \mid \mathbf{x}_{t}) = \text{Poisson}(\mby_{t_g,t_r}^s; \mathbf{I}_t).
    \end{eqnarray}
In our experiments we consider two streams of measurements $\mbY_t = (\mby_{t_g,t_r}^1,\mby_{t_g,t_r}^2)$. The first stream of measurements has no delay and is taken at every time-step such that $t = t_g = t_r, \forall\mby^1, \forall t$. The second stream of measurements has a delay of $3$ and is taken every third time-step such that $t = t_g = t_r+3, \forall\mby^2, t=1,4,7,\dots$. Fig.~\ref{fig:flpf_images} shows a time series of graphical representation of the two measurement streams. The measurement denoted by a circle is obtained daily while the triangle is a measurement obtained with a latency of three days. 

An example of quantifying the number of observations processed by PFs with different lags, we accounted for both the daily (on-time) and OOS measurements. For a time series spanning 730 days, with one daily measurement per day and an additional OOS measurement available every three days, the PF with lag = 0 uses only the daily measurements (730 total). The PF with lag = 3 incorporates both the daily data and OOS measurements that arrive three days late, resulting in 242 OOS measurements. The PF with lag = 7 further includes OOS measurements that are delayed by six days, adding another 241 OOS measurements.

\subsection{Design of the performance analysis}

To assess the accuracy of the algorithms, we calculate the Mean Square Error (MSE) between the true outbreaks and the algorithm's predictions. Additionally, Receiver Operating Characteristic (ROC) curves are employed to evaluate the sensitivity of the algorithms to false alarms and Activity Monitoring Operator Characteristic (AMOC) curves illustrate the timeliness of detecting an outbreak.

\subsubsection{Mean Square Error}
\begin{figure} [!ht]
    \centering
    \includegraphics[width=1.0\linewidth]{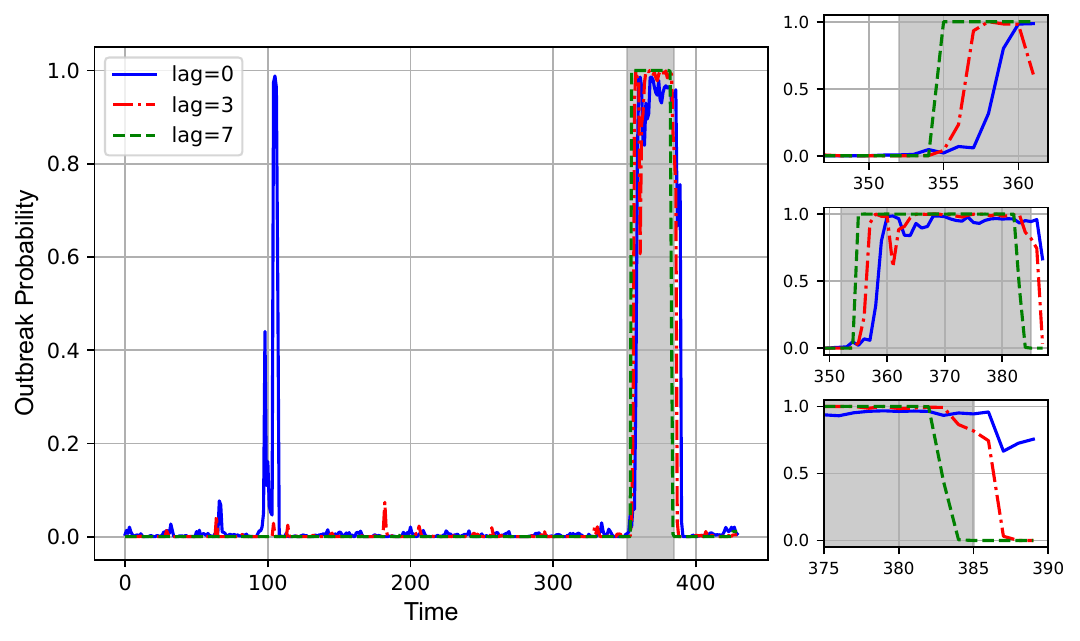}
    \caption{{\bf Estimates of the probability of outbreak with different lags in the PF.} Estimated outbreak probabilities compared to the true underlying outbreak. The gray shaded regions indicate the true outbreak periods, while colored lines show PF estimates with lag = 0 (solid blue), lag = 3 (red dash-dot), and lag = 7 (green dashed). The three inset panels highlight the onset, peak, and decline phases of the outbreak, illustrating that larger lags enable closer tracking of the true dynamics across all stages.} \label{fig:MSE_example}
\end{figure}

The MSE is a commonly used metric to evaluate the accuracy of a model. It measures the average of the squares of the errors, i.e., the average squared difference between the predicted values and the actual (true) values. The MSE gives a higher weight to larger errors, which makes it more sensitive to outliers. In this paper the MSE is used to compare the true probability of an outbreak and the estimate from the PF based on the evolving state of the system. This provides a quantitative measure of how well the filter is estimating each disease state. An example when estimating the probability of outbreak is defined below where for each time step $t$, you have:
\begin{itemize}
    \item $O_{true}(t)$ the actual outbreak indicator (1 if there's an outbreak, 0 if not),
    \item $P_{estimated}(t)$ the probability of an outbreak estimated by the PF.
\end{itemize}
The MSE for the outbreak probability can be computed as:
\begin{eqnarray}\label{eq:MSE_outbreak}
    \text{MSE}_{outbreak} = \frac{1}{T} \sum_{t=1}^{T} \left( O_{true}(t) - P_{estimated}(t) \right)^2,
\end{eqnarray}
where $T$ is the total number of time-steps. The insets in Fig.~\ref{fig:MSE_example} highlight the onset, peak, and decline phases of the outbreak, showing that higher lags (particularly lag = 7) more closely track the true outbreak probability across all stages, whereas smaller lags (lag = 0) tend to underestimate the probability during critical transition periods. This demonstrates how the MSE captures the cumulative deviation between the estimated and true outbreak probabilities over time and across different filter configurations.

\subsubsection{Receiver Operating Characteristic}
Making predictions regarding if there is an outbreak or not can be evaluated using these metrics:
\begin{itemize}
  \item Sensitivity (True Positive Rate): measures the proportion of actual outbreaks that the detection system correctly identifies.
  \item Specificity (True Negative Rate): measures the proportion of non-outbreak periods that the detection system correctly identifies as not being an outbreak.
  \item False alarm rate: measures the proportion of non-outbreak periods that the system incorrectly identifies as outbreaks.
\end{itemize}
The predicted probabilities of an outbreak from the PF and the true probabilities, as outlined in Eq~\ref{eq:MSE_outbreak}), can be used within Receiver Operating Characteristic (ROC) curves to provide a clear assessment of how well the predicted probabilities align with the actual outbreak events. The true status of an outbreak at time $t$, $0_{true}(t)$, can either be $0$ or $1$, depending if it is a non-outbreak or outbreak, respectively. The estimated probabilities from a PF can be a continuous value between $0$ and $1$, $P_{estimated}(t)\in\left[0,1\right]$, which represents the likelihood of an outbreak at time $t$. As the predicted probability from the PF is continuous, a threshold to the probabilities is applied to create a binary decision on whether an outbreak is predicted. For example
\begin{itemize}
    \item if $P_{estimated}(t)>\text{Threshold}$, an outbreak is predicted,
    \item if $P_{estimated}(t)\leq\text{Threshold}$, an outbreak is not predicted.
\end{itemize}
By varying the threshold from 0 to 1, we can evaluate the model's performance at different sensitivity-specificity trade-offs and generate the ROC curve. A ROC curve allows for a visualisation of the trade-offs between the True Positive Rate and the False Positive Rate.

\subsubsection{Activity Monitoring Operator Characteristic}
An AMOC curve is a graphical tool that illustrates the timeliness of a detection algorithm. It plots the expected time to detection against the false positive rate (or the number of false alarms), providing a mechanism to compare the performance of different detection algorithms. An AMOC curve depicts how quickly a detector identifies an outbreak relative to the rate of false alarms it generates. Assessing performance using an AMOC curve involves plotting the number of days to detection against the number of false alarms across various threshold values. This visualises the trade-off between early detection and the occurrence of false positives, giving insights into how effective the detector is in balancing timeliness with accuracy. Lets denote the time to detection (which can be measured in hours or days) as $T_D$ and expected time to detection, which represents the average delay to detection over multiple runs of the algorithm, as $\mathbb{E}[T_D]$. The time to detection can be measured as
    \begin{equation} 
        T_D = \text{min}\{t\mid D_{t}=1\text{ for some Threshold}\}.
    \end{equation}
where $D_t$ is a binary decision function that takes the value $1$ if the detector raises an alarm at time $t$ and $0$ otherwise. The AMOC curve is created by plotting $\mathbb{E}[T_D]$ against the false positive rate for various threshold values. 

\section{Results}
The analysis conducted in this paper was run on $1$ Distributed Memory (DM) machine. which consists of a `2 Xeon Gold 6138' CPU, which provides a memory of $384$GB and $40$ cores. The experiments requested a power-of-two number of cores and treated the cores as DM processors (i.e., $P = 32$). 

\subsection{Parameter estimation}

For parameter estimation, the setup consisted of a population of 10,000 individuals over a time series of two years (730 days) with the outbreak period between days 240-480. The true values for the parameters in the data generation were $\bm{\theta} =\lbrack\beta, \beta_1, \gamma, \sigma, \delta\rbrack = \lbrack 0.1, 0.3, 0.05, 0.08, 0.005\rbrack$. The PF was configured with $N_x=1024$ particles and and the PF transition matrix for switching between outbreak regimes was defined as 
    \begin{eqnarray}
        \mathcal{M} = \begin{bmatrix}
        0.999 & 0.001 \\
        0.011 & 0.989,
    \end{bmatrix}
    \end{eqnarray}
corresponding to probabilities of remaining in or transitioning between the no-outbreak and outbreak states. The configuration of the SMC sampler consisted of $K=50$ iterations, $N=1024$ samples, using the forwards proposal L-kernel and stepsizes of $\bm{\theta} =\lbrack\beta, \beta_1, \gamma, \sigma, \delta\rbrack = \lbrack 1 \times 10^{-4}, 1 \times 10^{-4}, 1 \times 10^{-4}, 1 \times 10^{-4}, 1 \times 10^{-6}\rbrack$. 

Before evaluating the ability of the PF to estimate the outbreak states, we first examine its ability to recover the model parameters. This step is crucial, as the proposed FL-PF framework is designed to be compatible with parameter estimation with SMC$^2$, allowing joint parameter and state inference.

\begin{figure*}[t]
\centering

\begin{subfigure}[t]{0.3\linewidth}
    \centering
    \includegraphics[width=\linewidth]{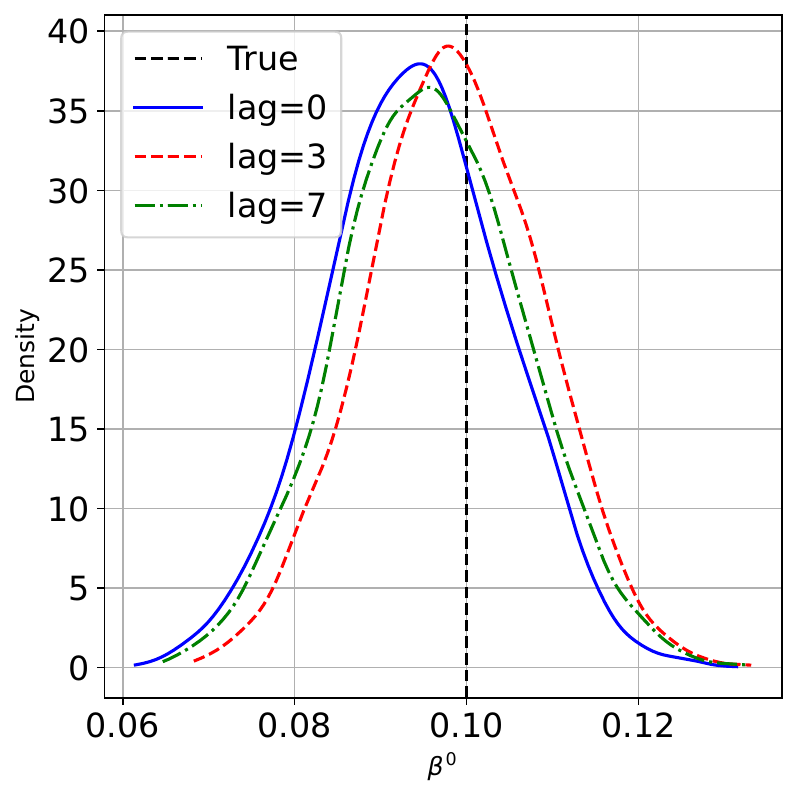}
    \caption{}
\end{subfigure}
\hfill
\begin{subfigure}[t]{0.3\linewidth}
    \centering
    \includegraphics[width=\linewidth]{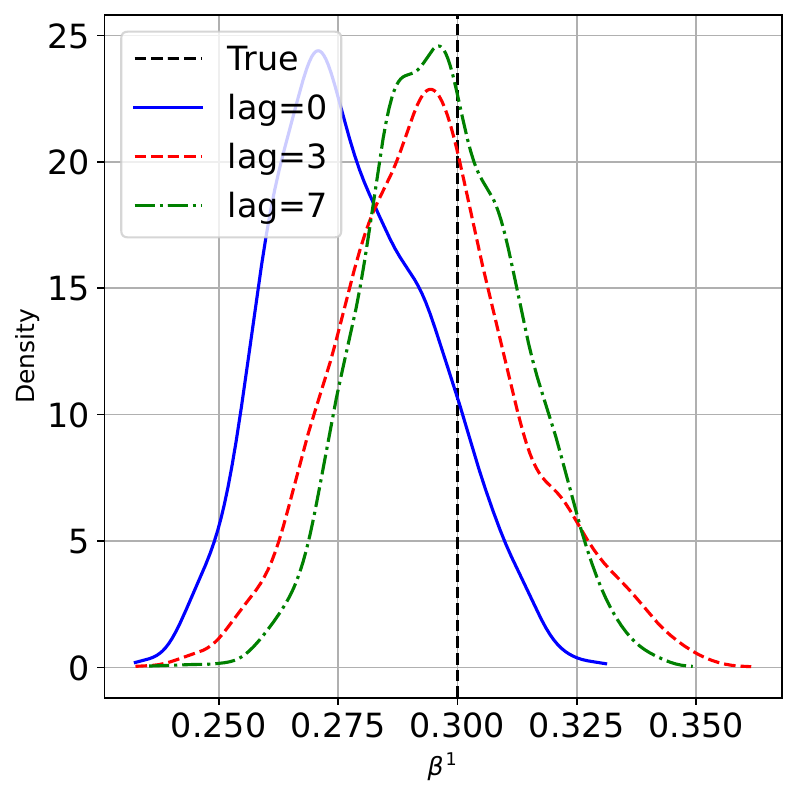}
    \caption{}
\end{subfigure}
\hfill
\begin{subfigure}[t]{0.3\linewidth}
    \centering
    \includegraphics[width=\linewidth]{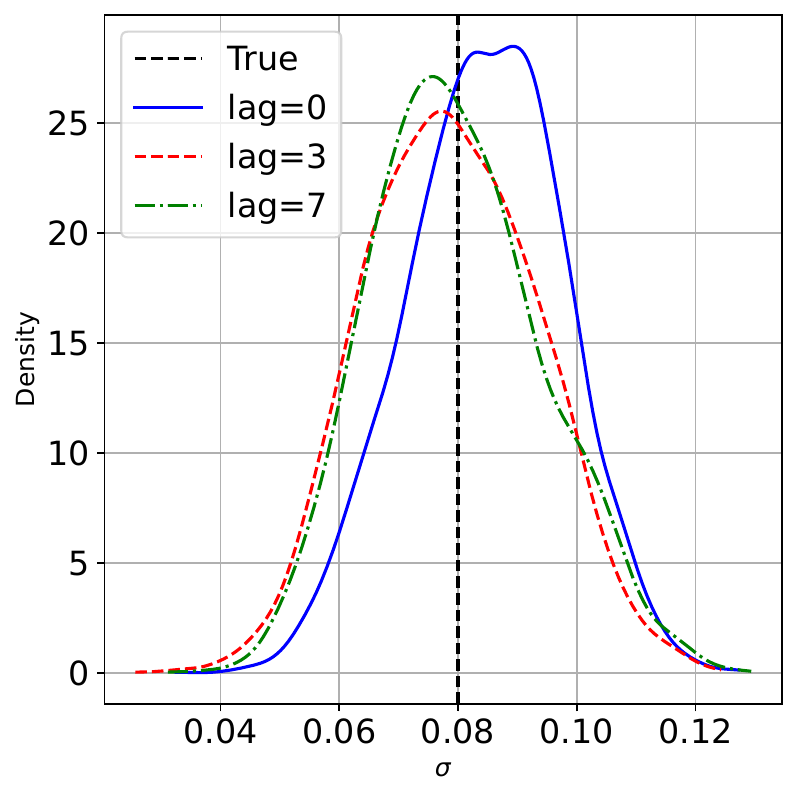}
    \caption{}
\end{subfigure}

\vspace{1em}

\begin{subfigure}[t]{0.3\linewidth}
    \centering
    \includegraphics[width=\linewidth]{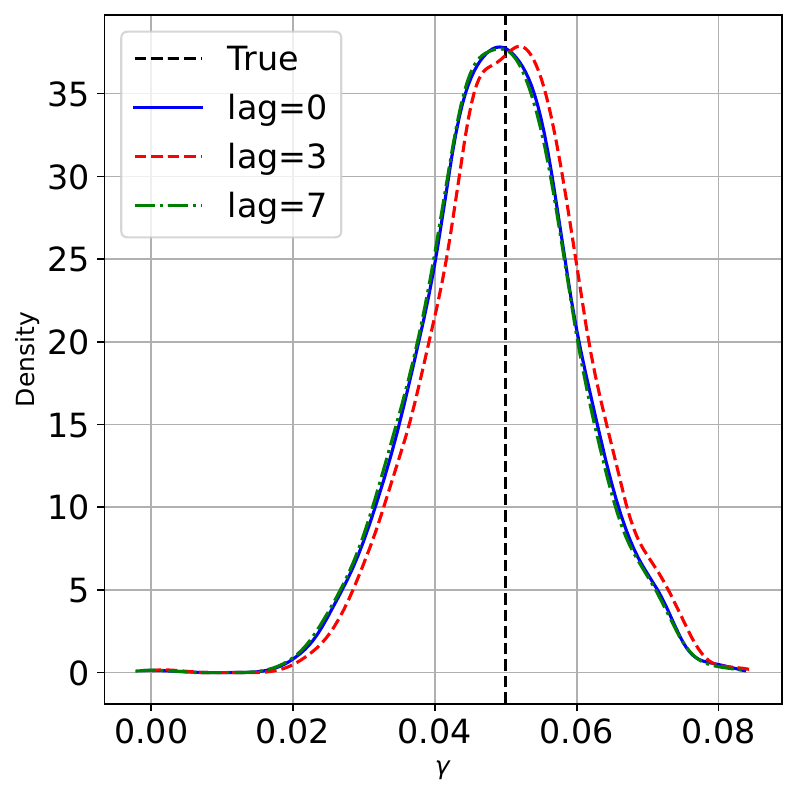}
    \caption{}
\end{subfigure}
\hfill
\begin{subfigure}[t]{0.3\linewidth}
    \centering
    \includegraphics[width=\linewidth]{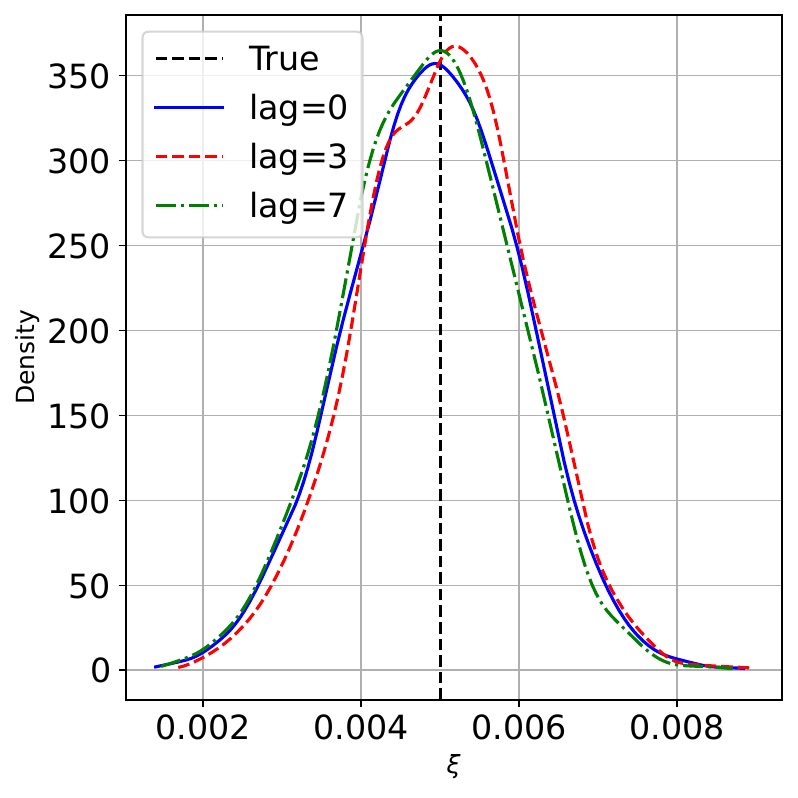}
    \caption{}
\end{subfigure}
\caption{{\bf Parameter estimation performance of the PF with different lags.} In all panels, the solid blue line denotes lag = 0, the red dash–dot line denotes lag = 3, and the green dotted line denotes lag = 7. The vertical black line indicates the true value.}
\label{fig:parameter_estimation}
\end{figure*}

Fig.~\ref{fig:parameter_estimation} shows the posterior distributions of the parameters estimated using the PF with different lags. The distributions are obtained from the final iteration of SMC samples across five random seeds. Across all lag settings, the parameter estimates closely match the true values, indicating that the PF accurately identifies the underlying disease dynamics even when the exact parameters are unknown.

Interestingly, the parameter estimates are nearly identical across different lags. This consistency suggests that the lag primarily affects the state estimation performance, which is explored in Section~\ref{sec:state_estimation}, rather than parameter estimation. In other words, while the parameters are robustly estimated in all cases, introducing a lag improves the quality of the outbreak trajectory estimates that rely on those parameters, as outlined in Section~\ref{sec:state_estimation}.


\subsection{State estimation}\label{sec:state_estimation}

For the data generation, we simulated a population of 10,000 individuals over a time series of two years (730 days) The true values for the parameters in the data generation were $\bm{\theta} = [\beta^0, \beta^1, \gamma, \sigma, \delta] = [0.2, 0.4, 1/5, 1/10, 1/180]$. The PF was configured with $N_x=512$ particles, and the PF transition matrix for switching between outbreak regimes was defined as 
    \begin{eqnarray}
        \mathcal{M} = \begin{bmatrix}
        0.999 & 0.001 \\
        0.011 & 0.989,
    \end{bmatrix}
    \end{eqnarray}
corresponding to probabilities of remaining in or transitioning between the no-outbreak and outbreak states. To evaluate accuracy, the MSE between the true and estimated outbreak probabilities was calculated starting from time $t=430$, ensuring that the epidemiological model had reached its steady state.

The performance metrics in Table~\ref{table1} indicate that incorporating a lag within the PF improves the ability to detect simulated outbreaks when using OOS measurements. While the zero-lag setting produces moderate performance across both single- and double-outbreak scenarios, introducing a lag of three leads to improvements, particularly in terms of reducing MSE and increasing the AUROC. The best overall performance is achieved at lag seven, which consistently produces the lowest MSE and AUAMOC values and the highest AUROC across both scenarios.

A key finding is that the standard deviation also decreases consistently with increasing lag. At a lag of seven, the standard deviations for all performance metrics are the lowest, indicating that the model's performance is not only better on average but also more reliable and consistent across different outbreak scenarios. This suggests that allowing the FL-PF to analyse delayed information helps to more accurately and reliably estimate outbreak probabilities, improving both sensitivity and stability in detection.

\begin{figure*}[]
\centering
\begin{subfigure}[t]{0.48\linewidth}
    \centering
    \includegraphics[width=\linewidth]{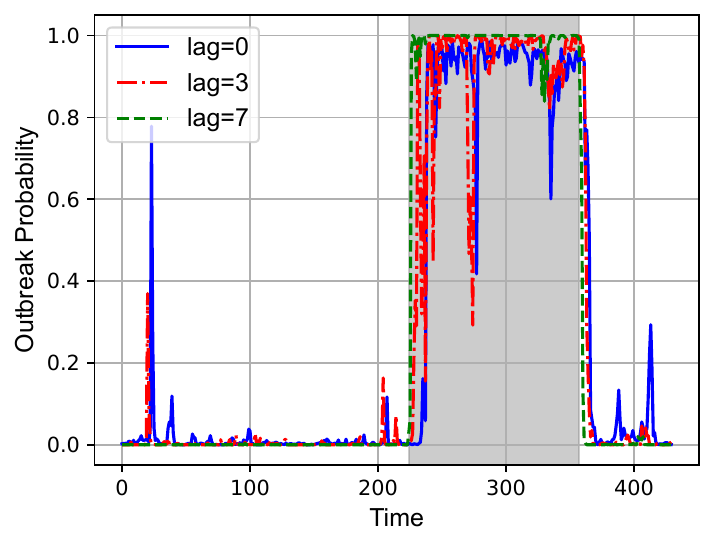}
    \caption{}
    \label{fig:roc_no_out_of_seq}
\end{subfigure}
\hfill
\begin{subfigure}[t]{0.48\linewidth}
    \centering
    \includegraphics[width=\linewidth]{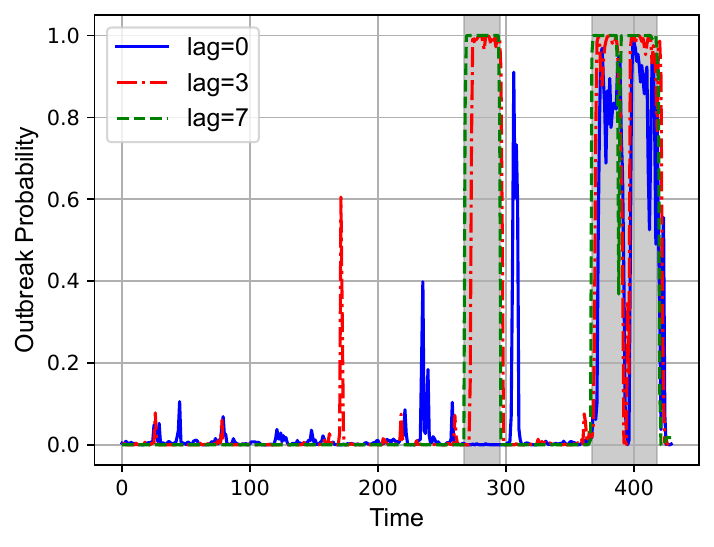}
    \caption{}
\end{subfigure}
\hfill
\begin{subfigure}[t]{0.48\linewidth}
    \centering
    \includegraphics[width=\linewidth]{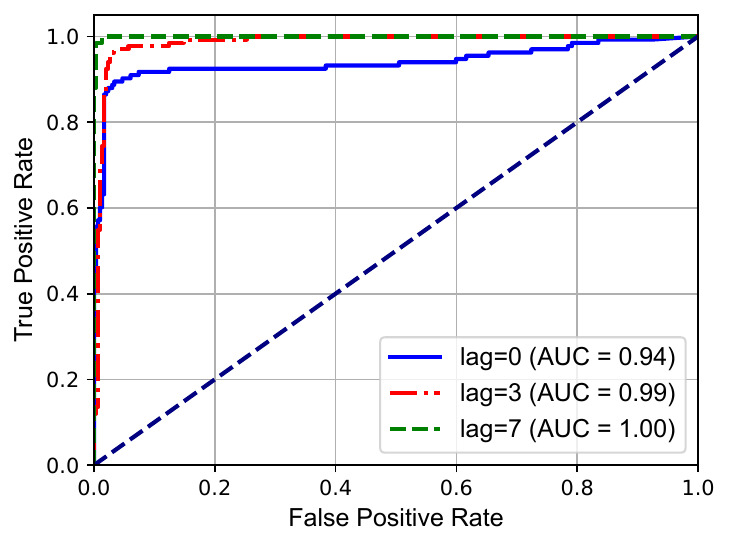}
    \caption{}
\end{subfigure}
\hfill
\begin{subfigure}[t]{0.48\linewidth}
    \centering
    \includegraphics[width=\linewidth]{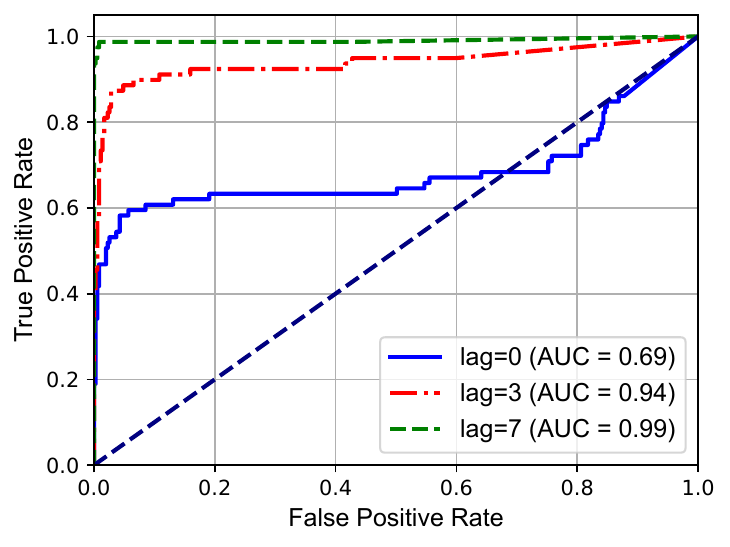}
    \caption{}
\end{subfigure}
\hfill
\begin{subfigure}[t]{0.48\linewidth}
    \centering
    \includegraphics[width=\linewidth]{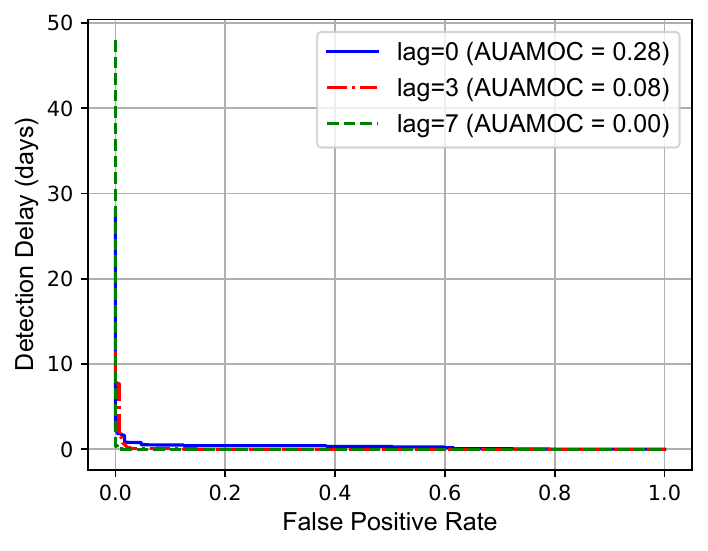}
    \caption{}
\end{subfigure}
\hfill
\begin{subfigure}[t]{0.48\linewidth}
    \centering
    \includegraphics[width=\linewidth]{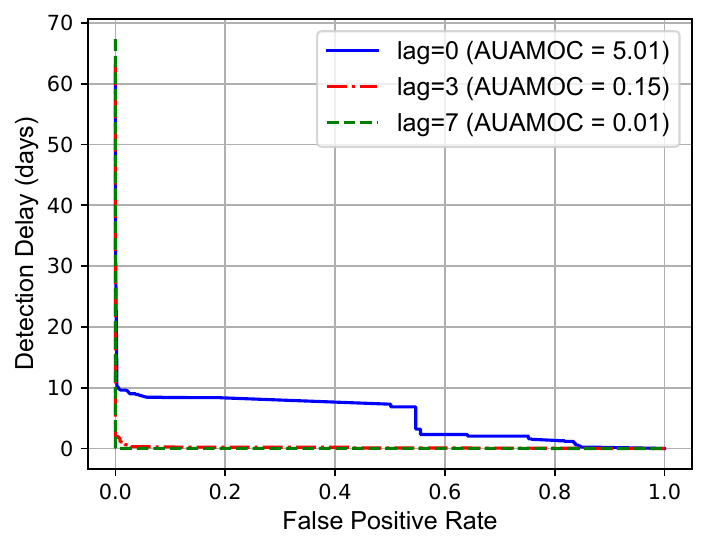}
    \caption{}
\end{subfigure}
\caption{{\bf State estimation performance of the PF with different lags.} The top row shows the predicted outbreak probability for (a) a single outbreak and (b) two outbreaks. Outbreaks correspond to the gray shaded time periods. The middle row presents the corresponding ROC curves, and the bottom row shows the AMOC curves. In all panels, the solid blue line denotes lag = 0, the red dash–dot line denotes lag = 3, and the green dotted line denotes lag = 7.}
\label{fig:state_estimation}
\end{figure*}

These trends are also illustrated in Fig.~\ref{fig:state_estimation}, which shows example predictions from simulations with one (panels A, C, E) and two outbreaks (panels B, D, F). The middle row highlights the improvement in discrimination ability through ROC curves, while the bottom row shows the corresponding AMOC curves, where smaller values indicate more efficient detection. Together, the quantitative results and visual examples demonstrate that lagged PFs are more robust in handling irregular data arrival, providing clearer and timelier identification of outbreak dynamics.

\begin{table}[!ht]
\centering
\small
\caption{
{\bf  Performance metrics for detecting one or two simulated outbreaks with OOS measurements and different lags within the PF. Results are averaged over 50 simulations containing one or two randomly generated outbreaks.} Results when estimating the probability of simulated outbreak(s) using the SEIRS model described in \protect{Section~\ref{sec:SEIRS_state}} from a PF with different lags. Two streams of data, one obtaining measurements daily and one with an OOS lag of three, are used. Both measurements are obtained from the Poisson likelihood outlined in \protect{Section~\ref{sec:SEIRS_likelihood}}. Smaller MSE and AUAMOC values are preferable, whereas a larger AUROC is desired. The bold value in each column outlines the best performing lag in the PF.}
\begin{tabular}{|l+l|l|l+l|l|l|l|l|l|l|}
\hline
& \multicolumn{3}{|c|}{\bf 1 Outbreak} & \multicolumn{3}{|c|}{\bf 2 Outbreaks} \\ \thickhline
 & MSE & AUROC & AUAMOC & MSE & AUAUROC & AUAMOC \\ \hline
Lag=0 & 0.049 (0.050)  & 0.808 (0.286) & 1.687(4.151) & 0.056 (0.030)  & 0.873 (0.173) & 1.042 (2.132) \\ \hline
Lag=3 & 0.027 (0.038)  & 0.926 (0.127) & 0.254 (0.850) & 0.036 (0.021) & 0.951 (0.053) & 0.162 (0.250)  \\ \hline
Lag=7 & \bf{0.011 (0.009)} & \bf{0.963 (0.112)} & \bf{0.164 (0.817)} & \bf{0.020 (0.011)} & \bf{0.967 (0.043)} & \bf{0.106 (0.249)}  \\ \hline
\end{tabular}
\begin{flushleft}

\end{flushleft}
\label{table1}
\end{table}

\section{Conclusion and future work}
The FL-PF is a valuable tool for disease modelling when dealing with delayed or OOS data. By explicitly incorporating the lagged observations into the filtering process, it provides more accurate state estimates. We have also outlined how to calculate the likelihood across different lags within a PF and provided empirical evidence that the parameter estimates using SMC$^2$ are consistent with the true values.

The disease model considered in this paper can be described as discrete such that the gradient of the log-likelihood with respect to $\bm{\theta}$ cannot be calculated. This limits the choice of proposal to the RW. Continuous representations of the SIR disease model, as outlined in \cite{rosato2022inference}, would allow for gradients to be calculated. This would allow more efficient proposals to be used, such as the Langevin proposal \cite{Langevin_proposal,murphy2025hess}, and the FL-PF with the No-U-Turn sampler (NUTS) \cite{NUTS} or change in estimator of the expected square (ChEES) \cite{millard2025incorporating} as the proposal \cite{FL_NUTS, acuto2025stone}.

The obvious extension of this work would be to apply the described methods to scenarios using real data pertinent to a disease. Using real data would require the SMC$^2$ algorithm to estimate the parameters of the model online. Therefore combining the methods outlined in this paper with the online SMC$^2$ algorithm described in \cite{TEMFACK2025100847} could be fruitful avenue for future work.

\bibliography{bibliography.bib}
\bibliographystyle{ieeetr}
\end{document}